\documentclass{article}

\usepackage{spconf}
\usepackage{amssymb, amsmath}
\usepackage{amsfonts}
\usepackage{bbm}
\usepackage{mathrsfs}
\usepackage{xspace}
\usepackage{bm}

\usepackage{cite}
\usepackage{epsfig}
\usepackage{cases}

\newtheorem{proposition}{Prop.}

\newenvironment{textbmatrix}{	\setlength{\arraycolsep}{2.5pt}%
								\big[\begin{matrix}}{\end{matrix}\big]%
								\raisebox{0.08ex}{\vphantom{M}}}

\def\be{\begin{equation}}
\def\ee{\end{equation}}
\def\een{\nonumber \end{equation}}
\def\mat{\begin{bmatrix}}
\def\emat{\end{bmatrix}}
\def\btm{\begin{textbmatrix}}
\def\etm{\end{textbmatrix}}

\def\ba#1\ea{\begin{align}#1\end{align}}
\def\bs#1\es{\begin{split}#1\end{split}} 
\def\bg#1\eg{\begin{gather}#1\end{gather}} 
\def\bi#1\ei{\begin{itemize}#1\end{itemize}}

\newcommand{\safemath}[2]{\newcommand{#1}{\ensuremath{#2}\xspace}}

\DeclareMathOperator{\diag}{diag}			
\DeclareMathOperator{\Prob}{\mathbb{P}}		

\safemath{\interior}{\mathrm{Int}}			 
\newcommand{\tp}[1]{\ensuremath{#1^{T}}} 		
\newcommand{\herm}[1]{\ensuremath{#1^{H}}} 	

\safemath{\dfn}{:=}							
\safemath{\dirac}{\delta}					

\safemath{\SNR}{\text{\sc snr}} 				
\safemath{\No}{N_0}							
\safemath{\Es}{E_s}							
\safemath{\Eb}{E_b}							
\safemath{\EbNo}{\frac{\Eb}{\No}}
\safemath{\EsNo}{\frac{\Es}{\No}}

\DeclareMathOperator{\CHop}{\ensuremath{\mathbb{H}}} 
\safemath{\tvir}{h_{\CHop}}					
\safemath{\tvtf}{L_{\CHop}}					
\safemath{\spf}{S_{\CHop}}						
\safemath{\bff}{H_{\CHop}}					

\safemath{\ircf}{R_{h}}						
\safemath{\scf}{R_{S}}						
\safemath{\tfcf}{R_{L}}						
\safemath{\bfcf}{R_{H}}						

\safemath{\mi}{I}							
\safemath{\capacity}{C}						

\safemath{\normal}{\mathcal{N}}				
\safemath{\circnorm}{\mathcal{CN}}			
\safemath{\mchain}{\leftrightarrow}			

\safemath{\dB}{\,\mathrm{dB}}
\safemath{\dBm}{\,\mathrm{dBm}}
\safemath{\Hz}{\,\mathrm{Hz}}
\safemath{\kHz}{\,\mathrm{kHz}}
\safemath{\MHz}{\,\mathrm{MHz}}
\safemath{\GHz}{\,\mathrm{GHz}}
\safemath{\s}{\,\mathrm{s}}
\safemath{\ms}{\,\mathrm{ms}}
\safemath{\mus}{\,\mathrm{\mu s}}
\safemath{\ns}{\,\mathrm{ns}}
\safemath{\meter}{\,\mathrm{m}}
\safemath{\mm}{\,\mathrm{mm}}
\safemath{\cm}{\,\mathrm{cm}}
\safemath{\m}{\,\mathrm{m}}
\safemath{\W}{\,\mathrm{W}}
\safemath{\J}{\,\mathrm{J}}
\safemath{\K}{\,\mathrm{K}}
\safemath{\bit}{\,\mathrm{bit}}

\safemath{\define}{\triangleq}			

\safemath{\equivalent}{\sim}
\safemath{\distas}{\sim}					

\safemath{\reals}{\mathbb{R}}
\safemath{\positivereals}{\mathbb{R}^{+}}
\safemath{\integers}{\mathbb{Z}}
\safemath{\posint}{\mathbb{Z}_{+}}
\safemath{\naturals}{\mathbb{N}}
\safemath{\complexset}{\mathbb{C}}

\safemath{\setA}{\mathcal{A}}
\safemath{\setB}{\mathcal{B}}
\safemath{\setC}{\mathcal{C}}
\safemath{\setD}{\mathcal{D}}
\safemath{\setE}{\mathcal{E}}
\safemath{\setF}{\mathcal{F}}
\safemath{\setG}{\mathcal{G}}
\safemath{\setH}{\mathcal{H}}
\safemath{\setI}{\mathcal{I}}
\safemath{\setJ}{\mathcal{J}}
\safemath{\setK}{\mathcal{K}}
\safemath{\setL}{\mathcal{L}}
\safemath{\setM}{\mathcal{M}}
\safemath{\setN}{\mathcal{N}}
\safemath{\setO}{\mathcal{O}}
\safemath{\setP}{\mathcal{P}}
\safemath{\setQ}{\mathcal{Q}}
\safemath{\setR}{\mathcal{R}}
\safemath{\setS}{\mathcal{S}}
\safemath{\setT}{\mathcal{T}}
\safemath{\setU}{\mathcal{U}}
\safemath{\setV}{\mathcal{V}}
\safemath{\setW}{\mathcal{W}}
\safemath{\setX}{\mathcal{X}}
\safemath{\setY}{\mathcal{Y}}
\safemath{\setZ}{\mathcal{Z}}
\safemath{\emptySet}{\varnothing}

\safemath{\bma}{\mathbf{a}}
\safemath{\bmb}{\mathbf{b}}
\safemath{\bmc}{\mathbf{c}}
\safemath{\bmd}{\mathbf{d}}
\safemath{\bme}{\mathbf{e}}
\safemath{\bmf}{\mathbf{f}}
\safemath{\bmg}{\mathbf{g}}
\safemath{\bmh}{\mathbf{h}}
\safemath{\bmi}{\mathbf{i}}
\safemath{\bmj}{\mathbf{j}}
\safemath{\bmk}{\mathbf{k}}
\safemath{\bml}{\mathbf{l}}
\safemath{\bmm}{\mathbf{m}}
\safemath{\bmn}{\mathbf{n}}
\safemath{\bmo}{\mathbf{o}}
\safemath{\bmp}{\mathbf{p}}
\safemath{\bmq}{\mathbf{q}}
\safemath{\bmr}{\mathbf{r}}
\safemath{\bms}{\mathbf{s}}
\safemath{\bmt}{\mathbf{t}}
\safemath{\bmu}{\mathbf{u}}
\safemath{\bmv}{\mathbf{v}}
\safemath{\bmw}{\mathbf{w}}
\safemath{\bmx}{\mathbf{x}}
\safemath{\bmy}{\mathbf{y}}
\safemath{\bmz}{\mathbf{z}}

\bmdefine{\biad}{a}
\bmdefine{\bibd}{b}
\bmdefine{\bicd}{c}
\bmdefine{\bidd}{d}
\bmdefine{\bied}{e}
\bmdefine{\bifd}{f}
\bmdefine{\bigd}{g}
\bmdefine{\bihd}{h}
\bmdefine{\biid}{i}
\bmdefine{\bijd}{j}
\bmdefine{\bikd}{k}
\bmdefine{\bild}{l}
\bmdefine{\bimd}{m}
\bmdefine{\bind}{n}
\bmdefine{\biod}{o}
\bmdefine{\bipd}{p}
\bmdefine{\biqd}{q}
\bmdefine{\bird}{r}
\bmdefine{\bisd}{s}
\bmdefine{\bitd}{t}
\bmdefine{\biud}{u}
\bmdefine{\bivd}{v}
\bmdefine{\biwd}{w}
\bmdefine{\bixd}{x}
\bmdefine{\biyd}{y}
\bmdefine{\bizd}{z}

\bmdefine{\bixid}{\xi}
\bmdefine{\bilambdad}{\lambda}
\bmdefine{\bimud}{\mu}
\bmdefine{\bithetad}{\theta}
\bmdefine{\biphid}{\phi}

\safemath{\bmia}{\biad}
\safemath{\bmib}{\bibd}
\safemath{\bmic}{\bicd}
\safemath{\bmid}{\bidd}
\safemath{\bmie}{\bied}
\safemath{\bmif}{\bifd}
\safemath{\bmig}{\bigd}
\safemath{\bmih}{\bihd}
\safemath{\bmii}{\biid}
\safemath{\bmij}{\bijd}
\safemath{\bmik}{\bikd}
\safemath{\bmil}{\bild}
\safemath{\bmim}{\bimd}
\safemath{\bmin}{\bind}
\safemath{\bmio}{\biod}
\safemath{\bmip}{\bipd}
\safemath{\bmiq}{\biqd}
\safemath{\bmir}{\bird}
\safemath{\bmis}{\bisd}
\safemath{\bmit}{\bitd}
\safemath{\bmiu}{\biud}
\safemath{\bmiv}{\bivd}
\safemath{\bmiw}{\biwd}
\safemath{\bmix}{\bixd}
\safemath{\bmiy}{\biyd}
\safemath{\bmiz}{\bizd}

\safemath{\bmxi}{\bixid}
\safemath{\bmlambda}{\bilambdad}
\safemath{\bmmu}{\bimud}
\safemath{\bmtheta}{\bithetad}
\safemath{\bmphi}{\biphid}

\safemath{\bA}{\mathbf{A}}
\safemath{\bB}{\mathbf{B}}
\safemath{\bC}{\mathbf{C}}
\safemath{\bD}{\mathbf{D}}
\safemath{\bE}{\mathbf{E}}
\safemath{\bF}{\mathbf{F}}
\safemath{\bG}{\mathbf{G}}
\safemath{\bH}{\mathbf{H}}
\safemath{\bI}{\mathbf{I}}
\safemath{\bJ}{\mathbf{J}}
\safemath{\bK}{\mathbf{K}}
\safemath{\bL}{\mathbf{L}}
\safemath{\bM}{\mathbf{M}}
\safemath{\bN}{\mathbf{N}}
\safemath{\bO}{\mathbf{O}}
\safemath{\bP}{\mathbf{P}}
\safemath{\bQ}{\mathbf{Q}}
\safemath{\bR}{\mathbf{R}}
\safemath{\bS}{\mathbf{S}}
\safemath{\bT}{\mathbf{T}}
\safemath{\bU}{\mathbf{U}}
\safemath{\bV}{\mathbf{V}}
\safemath{\bW}{\mathbf{W}}
\safemath{\bX}{\mathbf{X}}
\safemath{\bY}{\mathbf{Y}}
\safemath{\bZ}{\mathbf{Z}}

\safemath{\bZero}{\mathbf{0}}

\bmdefine{\biAd}{A}
\bmdefine{\biBd}{B}
\bmdefine{\biCd}{C}
\bmdefine{\biDd}{D}
\bmdefine{\biEd}{E}
\bmdefine{\biFd}{F}
\bmdefine{\biGd}{G}
\bmdefine{\biHd}{H}
\bmdefine{\biId}{I}
\bmdefine{\biJd}{J}
\bmdefine{\biKd}{K}
\bmdefine{\biLd}{L}
\bmdefine{\biMd}{M}
\bmdefine{\biOd}{N}
\bmdefine{\biPd}{O}
\bmdefine{\biQd}{P}
\bmdefine{\biRd}{R}
\bmdefine{\biSd}{S}
\bmdefine{\biTd}{T}
\bmdefine{\biUd}{U}
\bmdefine{\biVd}{V}
\bmdefine{\biWd}{W}
\bmdefine{\biXd}{X}
\bmdefine{\biYd}{Y}
\bmdefine{\biZd}{Z}

\bmdefine{\biDelta}{\Delta}
\bmdefine{\biLambda}{\Lambda}
\bmdefine{\biPhi}{\Phi}
\bmdefine{\biSigma}{\Sigma}
\bmdefine{\biOmega}{\Omega}
\bmdefine{\biTheta}{\Theta}

\safemath{\bimA}{\biAd}
\safemath{\bimB}{\biBd}
\safemath{\bimC}{\biCd}
\safemath{\bimD}{\biDd}
\safemath{\bimE}{\biEd}
\safemath{\bimF}{\biFd}
\safemath{\bimG}{\biGd}
\safemath{\bimH}{\biHd}
\safemath{\bimI}{\biId}
\safemath{\bimJ}{\biJd}
\safemath{\bimK}{\biKd}
\safemath{\bimL}{\biLd}
\safemath{\bimM}{\biMd}
\safemath{\bimN}{\biNd}
\safemath{\bimO}{\biOd}
\safemath{\bimP}{\biPd}
\safemath{\bimQ}{\biQd}
\safemath{\bimR}{\biRd}
\safemath{\bimS}{\biSd}
\safemath{\bimT}{\biTd}
\safemath{\bimU}{\biUd}
\safemath{\bimV}{\biVd}
\safemath{\bimW}{\biWd}
\safemath{\bimX}{\biXd}
\safemath{\bimY}{\biYd}
\safemath{\bimZ}{\biZd}

\safemath{\bDelta}{\bielta}
\safemath{\bLambda}{\biLambda}
\safemath{\bPhi}{\biPhi}
\safemath{\bSigma}{\biSigma}
\safemath{\bOmega}{\biOmega}
\safemath{\bTheta}{\biTheta}

\safemath{\veca}{\bma}
\safemath{\vecb}{\bmb}
\safemath{\vecc}{\bmc}
\safemath{\vecd}{\bmd}
\safemath{\vece}{\bme}
\safemath{\vecf}{\bmf}
\safemath{\vecg}{\bmg}
\safemath{\vech}{\bmh}
\safemath{\veci}{\bmi}
\safemath{\vecj}{\bmj}
\safemath{\veck}{\bmk}
\safemath{\vecl}{\bml}
\safemath{\vecm}{\bmm}
\safemath{\vecn}{\bmn}
\safemath{\veco}{\bmo}
\safemath{\vecp}{\bmp}
\safemath{\vecq}{\bmq}
\safemath{\vecr}{\bmr}
\safemath{\vecs}{\bms}
\safemath{\vect}{\bmt}
\safemath{\vecu}{\bmu}
\safemath{\vecv}{\bmv}
\safemath{\vecw}{\bmw}
\safemath{\vecx}{\bmx}
\safemath{\vecy}{\bmy}
\safemath{\vecz}{\bmz}
\safemath{\vecZero}{\bZero}

\safemath{\vecxi}{\bmxi}
\safemath{\veclambda}{\bmlambda}
\safemath{\vecmu}{\bmmu}
\safemath{\vectheta}{\bmtheta}
\safemath{\vecphi}{\bmphi}

\safemath{\matA}{\bA}
\safemath{\matB}{\bB}
\safemath{\matC}{\bC}
\safemath{\matD}{\bD}
\safemath{\matE}{\bE}
\safemath{\matF}{\bF}
\safemath{\matG}{\bG}
\safemath{\matH}{\bH}
\safemath{\matI}{\bI}
\safemath{\matJ}{\bJ}
\safemath{\matK}{\bK}
\safemath{\matL}{\bL}
\safemath{\matM}{\bM}
\safemath{\matN}{\bN}
\safemath{\matO}{\bO}
\safemath{\matP}{\bP}
\safemath{\matQ}{\bQ}
\safemath{\matR}{\bR}
\safemath{\matS}{\bS}
\safemath{\matT}{\bT}
\safemath{\matU}{\bU}
\safemath{\matV}{\bV}
\safemath{\matW}{\bW}
\safemath{\matX}{\bX}
\safemath{\matY}{\bY}
\safemath{\matZ}{\bZ}
\safemath{\matZero}{\bZero}

\safemath{\matDelta}{\bDelta}
\safemath{\matLambda}{\bLambda}
\safemath{\matPhi}{\bPhi}
\safemath{\matSigma}{\bSigma}
\safemath{\matOmega}{\bOmega}
\safemath{\matTheta}{\bTheta}

\safemath{\matIdentity}{\matI}

\newcommand{\utility[2]}{u_{#1}(#2)}
\newcommand{\utilityStar[1]}{u^*_{#1}}

\newcommand{\rate[1]}{R_{#1}}
\safemath{\infobits}{D}
\safemath{\totalbits}{M}
\newcommand{\power[1]}{p_{#1}}
\newcommand{\powerStar[1]}{p^*_{#1}}
\newcommand{\SINR[1]}{\gamma_{#1}}
\newcommand{\SINRStar[1]}{\gamma^*_{#1}}
\safemath{\SINRStarInf}{\overline{\SINR[]}^*}
\newcommand{\efficiencyFunction[1]}{f(#1)}
\newcommand{\efficiencyFunctionPrime[1]}{f^\prime(#1)}

\newcommand{\channelresponse[2]}{c_{#1}(#2)}
\safemath{\chiptime}{T_c}
\safemath{\srake}{\pathno_S}
\newcommand{\delay[1]}{\tau_{#1}}

\safemath{\SP}{\text{SP}}
\safemath{\SI}{\text{SI}}
\safemath{\MAI}{\text{MAI}}
\newcommand{\channelgain[1]}{h_{#1}}
\newcommand{\hSP[1]}{h_{#1}^{(\SP)}}
\newcommand{\hSI[1]}{h_{#1}^{(\SI)}}
\newcommand{\hMAI[1]}{h_{#1}^{(\MAI)}}
\safemath{\varnoise}{\sigma^2}
\newcommand{\vectornorm}[1]{\left|\left|{#1}\right|\right|}
\newcommand{\vecpathgain[1]}{\bmalpha_{#1}}
\newcommand{\vecrakecoeff[1]}{\bmbeta_{#1}}
\newcommand{\matpathgain[1]}{\matA_{#1}}
\newcommand{\matrakecoeff[1]}{\matB_{#1}}
\newcommand{\matCoeffHsi}{\matPhi}
\newcommand{\coeffHsi[1]}{\phi_{#1}}

\safemath{\game}{G}
\safemath{\userset}{\setK}

\newcommand{\pmax[1]}{\overline{p}_{#1}}
\newcommand{\powerset[1]}{P_{#1}}
\newcommand{\powervect[1]}{\vecp_{#1}}

\safemath{\powerTimesHsp}{q}
\safemath{\varq}{\sigma^2_\powerTimesHsp}
\safemath{\meanq}{\eta_\powerTimesHsp}

\safemath{\Po}{P_o}

\newcommand{\functionMu[1]}{\mu\left({#1}\right)}
\newcommand{\functionNu[1]}{\nu\left({#1}\right)}
\newcommand{\functionMuA[1]}{\mu_A\left({#1}\right)}
\newcommand{\functionNuA[1]}{\nu_A\left({#1}\right)}
\newcommand{\varUser[1]}{\sigma^2_{{#1}}}

\safemath{\PDPratio}{\Lambda}
\newcommand{\pathgain[2]}{\alpha_{#1}^{(#2)}}
\newcommand{\rakecoeff[2]}{\beta_{#1}^{(#2)}}
\safemath{\pathno}{L}
\safemath{\prake}{\pathno_P}
\safemath{\Pratio}{r}
\safemath{\userno}{K}
\safemath{\frameno}{N_f}
\safemath{\pulseno}{N_c}
\safemath{\gain}{N}
\safemath{\processingMatrix}{\bG}

\safemath{\loadFactor}{\rho}

\newcommand{\SIratio[1]}{\gamma_{0,{#1}}}
\newcommand{\MAIratio[1]}{\zeta_{#1}}
\safemath{\as}{\stackrel{a.s.}{\rightarrow}}
\safemath{\loss}{\Psi}
\newcommand{\distance[1]}{d_{#1}}

\newcounter{mytempeqncnt}

\title{Large System Analysis of Game-Theoretic Power Control in
UWB Wireless Networks with Rake Receivers}

\twoauthors
    {Giacomo Bacci, Marco Luise}
        {University of Pisa - Dip. Ing. Informazione\\
        Via Caruso - 56122 Pisa, Italy}
 {H. Vincent Poor\sthanks{This research
  was supported in part by the
  U. S. Air Force Research Laboratory under Cooperative
  Agreement No. FA8750-06-1-0252, and in part by the
  U. S. Defense Advanced Research Projects Agency
  under Grant HR0011-06-1-0052.}}
        {Princeton University - Dept. Electrical Eng.\\
        Olden Street - 08544 Princeton, NJ, USA}

\begin{document}

\maketitle

\begin{abstract}
This paper studies the performance of partial-Rake (PRake) receivers
in impulse-radio ultrawideband wireless networks when an energy-efficient
power control scheme is adopted. Due to the large bandwidth of the system,
the multipath channel is assumed to be frequency-selective. By using
noncooperative game-theoretic models and large system analysis,
explicit expressions are derived in terms of network parameters to measure the
effects of self- and multiple-access interference at a receiving
access point. Performance of the PRake is compared in terms of
achieved utilities and loss to that of the all-Rake receiver.
\end{abstract}

\section{Introduction}\label{sec:intro}

Ultrawideband (UWB) technology is considered to be a potential
candidate for next-generation multiuser data networks, due to its large
spreading factor (which implies large multiuser capacity) and its lower
spectral density (which allows coexistence with incumbent systems). 
The requirements for designing high-speed data mobile
terminals include efficient resource allocation and interference reduction.
These issues aim to allow each user to achieve the required quality of
service at the uplink receiver without causing unnecessary interference
to other users in the system, and minimizing power consumption.
Scalable energy-efficient power control (PC) techniques can be derived using
game theory \cite{saraydar2, bacci}. 

In this work, performance of partial Rake (PRake) receivers \cite{proakis} 
is studied in terms of transmit powers and utilities achieved in the
uplink of an infrastructure network at the Nash equilibrium, where utility
here is defined as the ratio of throughput to transmit power. By using the
large system analysis proposed in \cite{bacci}, we obtain a general
characterization for the terms due to self-interference (SI) and  multiple
access interference (MAI). Explicit expressions for the utilities achieved at 
the Nash equilibrium are then derived, and an approximation for the loss of 
PRake receivers with respect to (wrt) all-Rake (ARake) receivers is proposed.

The remainder of the paper is organized as follows. Some background for this
work is given in Sect. \ref{sec:background}, where the system model and the
results of the game-theoretic PC approach are shown.
In Sect. \ref{sec:interference}, large system analysis is used to
evaluate the effects of the interference at the Nash equilibrium. Performance
of the PRake at the Nash equilibrium is analyzed in
Sect. \ref{sec:performance}, where also a comparison with simulations
is provided. Some conclusions are drawn in Sect. \ref{sec:conclusion}.

\section{Background}\label{sec:background}

\subsection{System Model}\label{sec:model}

Commonly, impulse-radio (IR) systems are employed to implement UWB
systems. We focus here on a binary phase shift keying (BPSK)
time hopping (TH) IR-UWB system with
polarity randomization \cite{gezici1}. A network with \userno users 
transmitting to a common concentration point is
considered. The processing gain is 
$\gain=\frameno\cdot\pulseno$, where \frameno is the number of pulses that
represent one information symbol, and \pulseno is the number of possible
pulse positions in a frame \cite{gezici1}. The transmission is assumed to be
over \emph{frequency selective channels}, with the channel for user $k$
modeled as a tapped delay line:
\be
  \label{eq:channel}
  \channelresponse[k]{t} =
  \sum_{l=1}^{\pathno}{\pathgain[l]{k}\delta(t-(l-1)\chiptime-\delay[k])},
\ee
where \chiptime is the duration of the transmitted UWB pulse; \pathno
is the number of channel paths;
$\vecpathgain[k]=\tp{[\pathgain[1]{k},\dots,\pathgain[\pathno]{k}]}$ and
$\delay[k]$ are the fading coefficients and the delay of user $k$,
respectively. Considering a chip-synchronous scenario, the symbols are
misaligned by an integer multiple of \chiptime:
$\delay[k] = \Delta_k\chiptime$, for every $k$, where $\Delta_k$ is
uniformly distributed in $\{0,1,\dots,\gain-1\}$. In addition, we
assume that the channel characteristics remain unchanged over a
number of symbol intervals \cite{gezici1}.

Due to high resolution of UWB signals, multipath channels can have hundreds
of components, especially in indoor environments. To mitigate the
effect of multipath fading as much as possible, we consider an access point
where \userno Rake receivers\cite{proakis} are used.\footnote{For ease of
calculation, perfect channel estimation is considered throughout the paper.}
The Rake receiver for user $k$ is in general composed of \pathno coefficients,
where the vector $\vecrakecoeff[k]=\processingMatrix\cdot\vecpathgain[k]=
\tp{[\rakecoeff[1]{k},\dots,\rakecoeff[\pathno]{k}]}$ represents the combining
weights for user $k$, and the $\pathno\times\pathno$ matrix
$\processingMatrix$ depends on the type of Rake receiver employed.
In particular, if $\{\processingMatrix\}_{ll}=1$ for 
$1\le l\le\Pratio\cdot\pathno$, and $0$ elsewhere, 
where $\Pratio\triangleq\prake/\pathno$ and $0<\prake\le\pathno$, a
PRake with $\prake$ fingers using maximal ratio combining (MRC) scheme is
considered. It is worth noting that, when $\Pratio=1$, an ARake is implemented.

The signal-to-interference-plus-noise ratio (SINR) of the $k$th user
at the output of the Rake receiver can be well
approximated (for large \frameno, typically at least 5) by \cite{gezici1}
\be
  \label{eq:sinr}
  \SINR[k] = \frac{\hSP[k]\power[k]}{\displaystyle{\hSI[k]\power[k] +
      \sum_{j\neq k}{\hMAI[kj]\power[j]} +
      \sigma^2}},
\ee
where \varnoise is the variance of the additive white Gaussian
noise (AWGN) at the receiver; $\power[k]$ denotes the transmit power
of user $k$; and the gains are expressed by
\begin{align}
  \label{eq:hSP}
  \hSP[k] &= \herm{\vecrakecoeff[k]}\cdot\vecpathgain[k],\\
  \label{eq:hSI}
  \hSI[k] &= \frac{1}{\gain}
  \frac{\vectornorm{\matCoeffHsi\cdot
      \left(\herm{\matrakecoeff[k]}\cdot\vecpathgain[k]+
      \herm{\matpathgain[k]}\cdot\vecrakecoeff[k]\right)}^2}
       {\herm{\vecrakecoeff[k]}\cdot\vecpathgain[k]},\\
  \label{eq:hMAI}
  \hMAI[kj] &= \frac{1}{\gain}
  \frac{\vectornorm{\herm{\matrakecoeff[k]}\cdot\vecpathgain[j]}^2
  + \vectornorm{\herm{\matpathgain[j]}\cdot\vecrakecoeff[k]}^2
  + \left|\herm{\vecrakecoeff[k]}\cdot\vecpathgain[j]\right|^2}
  {\herm{\vecrakecoeff[k]}\cdot\vecpathgain[k]},
\end{align}
where the matrices
\begin{align}
  \label{eq:matrixA}
  \matpathgain[k] &=
  \begin{pmatrix}
    \pathgain[\pathno]{k}&\cdots&\cdots&\pathgain[2]{k}\\
    0&\pathgain[\pathno]{k}&\cdots&\pathgain[3]{k}\\
    \vdots&\ddots&\ddots&\vdots\\
    0&\cdots&0&\pathgain[\pathno]{k}\\
    0&\cdots&\cdots&0
  \end{pmatrix},\\
  \label{eq:matrixB}
  \matrakecoeff[k] &=
  \begin{pmatrix}
    \rakecoeff[\pathno]{k}&\cdots&\cdots&\rakecoeff[2]{k}\\
    0&\rakecoeff[\pathno]{k}&\cdots&\rakecoeff[3]{k}\\
    \vdots&\ddots&\ddots&\vdots\\
    0&\cdots&0&\rakecoeff[\pathno]{k}\\
    0&\cdots&\cdots&0
  \end{pmatrix},\\
  \label{eq:matrixPhi}
  \matCoeffHsi &=
  \diag\left\{\coeffHsi[1],\dots,\coeffHsi[\pathno-1]\right\},
  \quad \coeffHsi[l]=\sqrt{\tfrac{\min\{\pathno-l,\pulseno\}}{\pulseno}},
\end{align}
have been introduced for convenience of notation.

\subsection{The Game-Theoretic Power Control Game}\label{sec:npcg}

Consider the application of noncooperative PC techniques to the wireless
network described above. Focusing on mobile terminals, where it is often more
important to maximize the number of bits transmitted per Joule of energy
consumed than to maximize throughput, a game-theoretic energy-efficient 
approach as the one described in \cite{bacci} is considered.

We examine a noncooperative PC game in which each user seeks to maximize its 
own utility function as follows. Let $\game = [\userset, \{\powerset[k]\},
\{\utility[k]{\powervect[]}\}]$ be the proposed game where
$\userset=\{1,\dots,\userno\}$ is the index set for the users;
$\powerset[k]=[0, \pmax[]]$ is the strategy set, with
$\pmax[]$ denoting the maximum power constraint; 
and $\utility[k]{\powervect[]}$ is the payoff function for user 
$k$ \cite{saraydar2}:
\be
  \label{eq:utility}
  \utility[k]{\powervect[]}=\frac{\infobits}{\totalbits}\rate[]
  \frac{\efficiencyFunction[{\SINR[k]}]}{\power[k]},
\ee
where $\powervect[]=[\power[1],\dots,\power[\userno]]$ are the 
transmit powers; $\infobits$ is the number of information bits per 
packet; $\totalbits$ is the total number of bits per packet; $\rate[]$ 
is the transmission rate; $\SINR[k]$ is the SINR (\ref{eq:sinr}) for 
user $k$; and $\efficiencyFunction[{\SINR[k]}]$ is the efficiency function
representing the packet success rate (PSR), i.e., the probability that a
packet is received without an error. 

When the efficiency function is increasing, S-shaped \cite{saraydar2}, and
continuously differentiable, with $\efficiencyFunction[0]=0$,
$\efficiencyFunction[+\infty]=1$, $\efficiencyFunctionPrime[0]=
d\efficiencyFunction[{\SINR[k]}]/d\SINR[k]|_{\SINR[k]=0}=0$,
it has been shown \cite{bacci} that the solution of the maximization problem
$\max_{\power[k]\in\powerset[k]} \utility[k]{\powervect[]}$ is
\be
  \label{eq:powerStar}
  \powerStar[k]=\min\left\{
  \frac{\SINRStar[k]\left(\sum_{j\neq k}{\hMAI[kj]\power[j]}+
    \sigma^2\right)}{\hSP[k]\left(1-\SINRStar[k]/\SIratio[k]\right)},
  \pmax[]\right\},
\ee
where $\SINRStar[k]$ is the solution of
\be\label{eq:f_der}
  \efficiencyFunctionPrime[{\SINRStar[k]}]
  \SINRStar[k]\left(1-\SINRStar[k]/\SIratio[k]\right)=
  \efficiencyFunction[{\SINRStar[k]}],
\ee
with $\efficiencyFunctionPrime[{\SINRStar[k]}]=
d\efficiencyFunction[{\SINR[k]}]/d\SINR[k]|_{\SINR[k]=\SINRStar[k]}$, and
$\SIratio[k]=\hSP[k]/\hSI[k]$.

\begin{figure*}[!t]
  \normalsize
  \setcounter{mytempeqncnt}{\value{equation}}
  \setcounter{equation}{16}
  \be\label{eq:functionNu}
    \functionNu[{\PDPratio,\Pratio,\loadFactor}]=
    \begin{cases}
      \tfrac{
        \PDPratio\left(\PDPratio^{\loadFactor}-1\right)
        \left(4\PDPratio^{2\Pratio}+3\PDPratio^{\loadFactor}-1\right)
        -2\PDPratio^{\Pratio+\loadFactor}\left(\PDPratio^{\Pratio}+
        3\PDPratio-1\right)
        \loadFactor\log{\PDPratio}}
            {2\left(\PDPratio^{\Pratio}-1\right)^2\loadFactor
              \PDPratio^{1+\loadFactor}\log{\PDPratio}},
           &\text{if $0\le\loadFactor\le\min(\Pratio,1-\Pratio)$};\\
      \tfrac{
        \PDPratio\left(4\PDPratio^{\loadFactor}-1\right)
        \left(\PDPratio^{2\Pratio}-1\right)
        -2\PDPratio^{\Pratio+\loadFactor}
        \left(3\PDPratio\Pratio-\loadFactor+\PDPratio^{\Pratio}
        \loadFactor\right)
        \log{\PDPratio}}
            {2\left(\PDPratio^{\Pratio}-1\right)^2\loadFactor
              \PDPratio^{1+\loadFactor}\log{\PDPratio}},
           &\text{if $\Pratio\le\loadFactor\le1-\Pratio$
             and}\\
           &\,\,\,\,\,\,\text{$\Pratio\le1/2$};\\
      \tfrac{
        -4\PDPratio^{2+2\Pratio}-4\PDPratio^{2+\loadFactor}+
          \PDPratio^{2(\Pratio+\loadFactor)}+4
          \PDPratio^{2+2\Pratio+\loadFactor}+
          3\PDPratio^{2+2\loadFactor}-2\PDPratio^{1+\Pratio+\loadFactor}
          \left(\Pratio+3\PDPratio\loadFactor+\PDPratio^{\Pratio}
          \loadFactor-1\right)
          \log{\PDPratio}}
           {2\left(\PDPratio^{\Pratio}-1\right)^2\loadFactor
             \PDPratio^{2+\loadFactor}\log{\PDPratio}},
           &\text{if $1-\Pratio\le\loadFactor\le\Pratio$
             and}\\
           &\,\,\,\,\,\,\text{$\Pratio\ge1/2$};\\
      \tfrac{
        -\PDPratio^{2+2\Pratio}-4\PDPratio^{2+\loadFactor}+
          \PDPratio^{2(\Pratio+\loadFactor)}+
          4\PDPratio^{2+2\Pratio+\loadFactor}-
          2\PDPratio^{1+\Pratio+\loadFactor}
          \left(\Pratio+3\PDPratio\loadFactor+
          \PDPratio^{\Pratio}\loadFactor-1\right)
          \log{\PDPratio}}
           {2\left(\PDPratio^{\Pratio}-1\right)^2
             \loadFactor\PDPratio^{2+\loadFactor}\log{\PDPratio}}
           &\text{if $\max(\Pratio,1-\Pratio)\le\loadFactor\le1$};\\
      \tfrac{2\PDPratio\left(\PDPratio^{2\Pratio}-1\right)-
        \left(\PDPratio^{\Pratio}+\Pratio+3\PDPratio\Pratio-1\right)
        \PDPratio^{\Pratio}\log{\PDPratio}}
           {\left(\PDPratio^{\Pratio}-1\right)^2
             \loadFactor\PDPratio\log{\PDPratio}},
           &\text{if $\loadFactor\ge1$}.
    \end{cases}
  \ee
  \setcounter{equation}{\value{mytempeqncnt}}
  \hrulefill
  \vspace*{4pt}
\end{figure*}

In the typical case of multiuser UWB systems, $\gain\gg\userno$.
If $\pmax[]$ is sufficiently large, (\ref{eq:powerStar}) can be reduced
to \cite{bacci}
\be
  \label{eq:minimumPower}
  \powerStar[k]=\frac{1}{\hSP[k]}\cdot
  \frac{\sigma^2\SINRStarInf}
       {1-\SINRStarInf\cdot
         \left(\SIratio[k]^{-1}+\MAIratio[k]^{-1}\right)},
\ee
where $\MAIratio[k]^{-1}=\sum_{j\neq k}{\hMAI[kj]/\hSP[j]}$;
and \SINRStarInf is the SINR at the Nash equilibrium in the absence of SI,
i.e., it is the solution of (\ref{eq:f_der}) when $\SIratio[k]=+\infty$.

A necessary and sufficient condition for the Nash equilibrium to be
achieved simultaneously by all the \userno users, and thus for
(\ref{eq:minimumPower}) to be valid, is \cite{bacci}
\be
  \label{eq:requirement}
  \SINRStarInf\cdot\left(\SIratio[k]^{-1}+\MAIratio[k]^{-1}\right)<1
  \quad \forall k\in\userset.
\ee

\section{Analysis of the Interference}\label{sec:interference}

\subsection{Analytical Results}

As can be verified in (\ref{eq:minimumPower}), the amount of transmit 
power $\powerStar[k]$ required
to achieve the target SINR $\SINRStar[k]$ will depend not only on the gain
$\hSP[k]$, but also on the SI term $\hSI[k]$ (through $\SIratio[k]$) and
the interferers $\hMAI[kj]$ (through $\MAIratio[k]$).
To derive quantitative results for the transmit powers independent of SI 
and MAI terms, it is possible to resort to a large-system 
analysis \cite{bacci}.

For ease of calculation, the expressions derived in the remainder of the paper
consider the following assumptions:

\begin{itemize}
  \item The channel gains are assumed to be independent complex Gaussian
    random variables with zero mean and variance $\varUser[{k_l}]$, i.e.,
    $\pathgain[k]{l}\distas\circnorm(0, \varUser[{k_l}])$. This assumption
    leads $|\pathgain[k]{l}|$ to be Rayleigh-distributed with
    parameter $\varUser[{k_l}]/2$. Although channel modeling for UWB
    systems is still an open issue, this hypothesis, appealing for its
    analytical tractability, also provides a good approximation for multipath
    propagation in UWB systems \cite{schuster}.

  \item The averaged power delay profile (aPDP) is
    assumed to decay exponentially, as is customarily taken in most UWB
    channel models \cite{molisch}. Hence, 
    $\varUser[{k_l}]=\varUser[k]\cdot\PDPratio^{-\frac{l-1}{\pathno-1}}$,
    where $\PDPratio=\varUser[{k_1}]/\varUser[{k_\pathno}]$ and
    $\varUser[k]$ depends on the distance between user
    $k$ and the base station. It is easy to verify that
    $\PDPratio=0\,\text{dB}$ represents the case of flat aPDP.
\end{itemize}

\begin{proposition}\label{th:new}
In the asymptotic case where \userno and \frameno are finite, while
$\pathno, \pulseno \rightarrow \infty$, when adopting a PRake with $\prake$
coefficients according to the MRC scheme, the terms $\MAIratio[k]^{-1}$ and
$\SIratio[k]^{-1}$ converge almost surely (a.s.) to
\begin{align}
  \label{eq:thMAI}
  \MAIratio[k]^{-1}&\as\frac{\userno-1}{\gain}\cdot
  \functionMu[{\PDPratio,\Pratio}],\\
  \label{eq:thSI}
  \SIratio[k]^{-1}&\as\frac{1}{\gain}\cdot
  \functionNu[{\PDPratio,\Pratio,\loadFactor}],\\
  \intertext{where $\Pratio\triangleq\prake/\pathno$, $0<\Pratio\le1$,
    and $\loadFactor\triangleq\pulseno/\pathno$, $0<\loadFactor<\infty$,
    are held constant, and}
  \label{eq:functionMu}
  \functionMu[{\PDPratio,\Pratio}]&=
  \frac{\left(\PDPratio-1\right)\cdot\PDPratio^{\Pratio-1}}
       {\PDPratio^{\Pratio}-1},
\end{align}
with $\functionNu[{\PDPratio,\Pratio,\loadFactor}]$ defined as in
(17), shown at the top of the page.
\end{proposition}

The proof of Prop. \ref{th:new} has been omitted because of space
limitation. It can be found in \cite{bacci2}.

Proposition \ref{th:new} gives accurate approximations for the terms of
MAI and SI in the case of PRake receivers at the access point and of
exponentially decaying aPDP. Results for more specific scenarios can be
derived using particular values of $\PDPratio$ and $\Pratio$, as shown in 
\cite{bacci2}. As an example, it is possible to obtain approximations for 
the MAI and SI arising in the ARake as follows:
\begin{align}
  \setcounter{equation}{17}
  \label{eq:muArakeExp}
  &\functionMuA[\PDPratio]=
  \lim_{\Pratio\to1}\functionMu[{\PDPratio,\Pratio}]=1,\\
  \label{eq:nuArakeExp}
  &\functionNuA[{\PDPratio,\loadFactor}]=
  \lim_{\Pratio\to1}\functionNu[{\PDPratio,\Pratio,\loadFactor}]=\nonumber\\
  &\quad=
  \begin{cases}
    \displaystyle{\frac{2\left(\PDPratio^2-1+\PDPratio^{\loadFactor}-
        \PDPratio^{2-\loadFactor}-
      2\PDPratio\loadFactor\log{\PDPratio}\right)}
         {\left(\PDPratio-1\right)^2\loadFactor\log{\PDPratio}}},
         & \text{if $\loadFactor\le1$},\\
    \displaystyle{\frac{2\left(\PDPratio^2-1-2\PDPratio\log{\PDPratio}\right)}
         {\left(\PDPratio-1\right)^2\loadFactor\log{\PDPratio}}},
         & \text{if $\loadFactor\ge1$}.
  \end{cases}
\end{align}

\subsection{Comments on the Results}

\begin{figure}
  \centering
  \includegraphics[width=8.5cm]{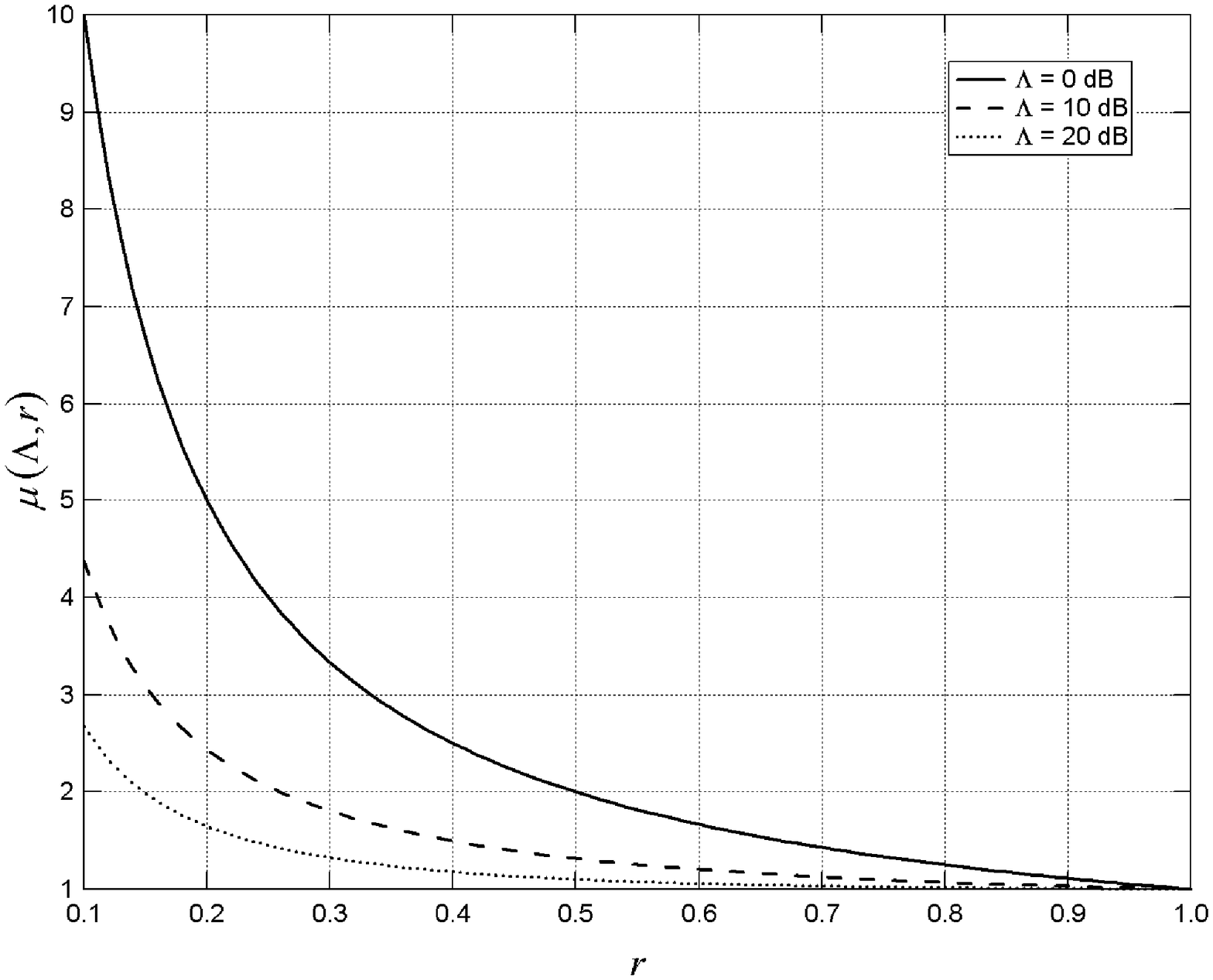}
  \caption{Shape of $\functionMu[{\PDPratio, \Pratio}]$ versus
    $\Pratio$ for some $\PDPratio$'s.}
  \label{fig:mu}
\end{figure}

Fig. \ref{fig:mu} shows the shape of $\functionMu[{\PDPratio, \Pratio}]$
versus $\Pratio$ for some values of $\PDPratio$. As can be noticed,
$\functionMu[{\PDPratio, \Pratio}]$ is decreasing as either $\PDPratio$ or
$\Pratio$ increases. Keeping $\Pratio$ fixed,
$\functionMu[{\PDPratio, \Pratio}]$ is a decreasing function of $\PDPratio$,
since the neglected paths are weaker as $\PDPratio$ increases. Keeping
$\PDPratio$ fixed, $\functionMu[{\PDPratio, \Pratio}]$ is a decreasing
function of $\Pratio$, since the receiver uses a higher number of
coefficients, thus better mitigating the effect of MAI. 

\begin{figure}
  \centering
  \includegraphics[width=8.5cm]{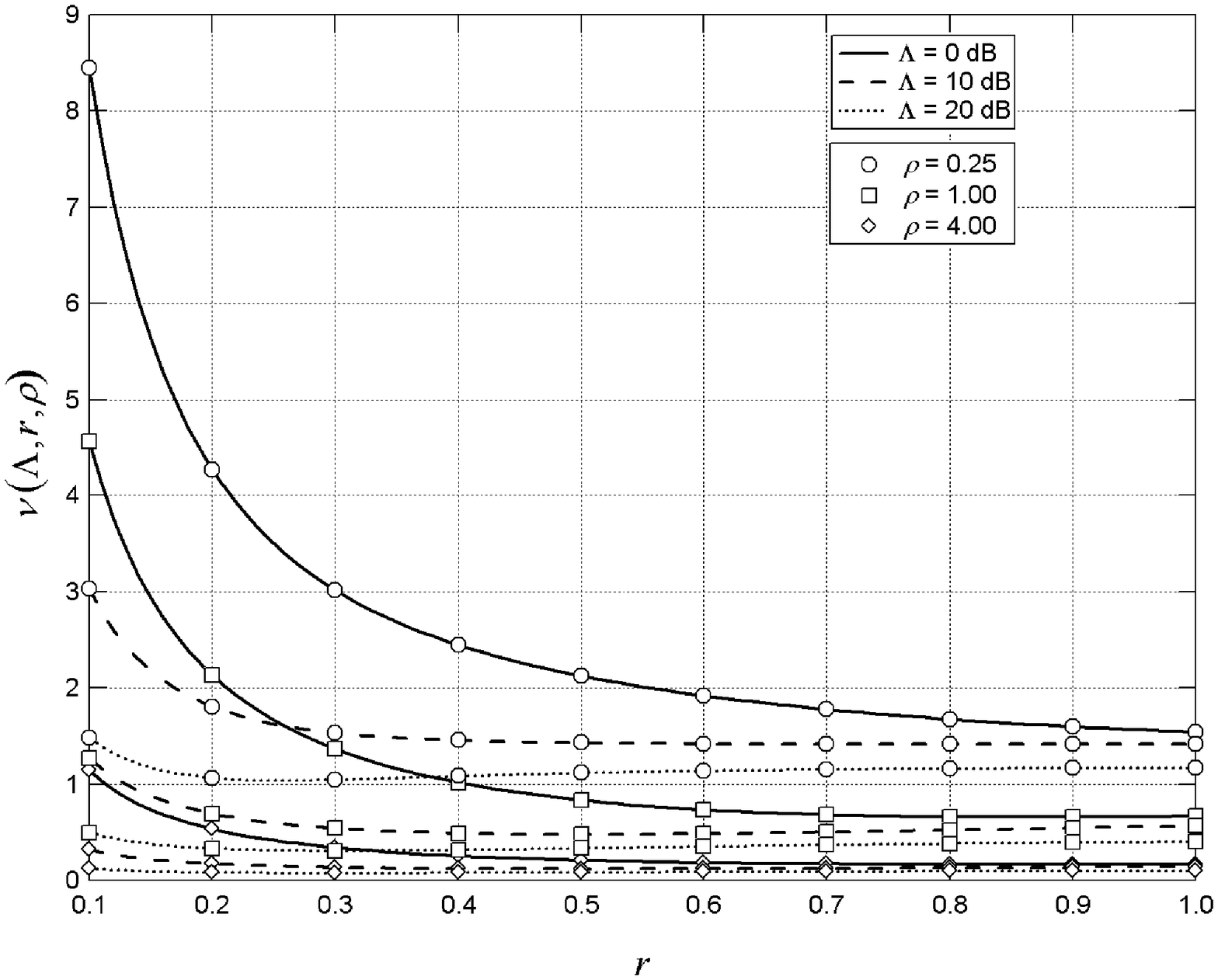}
  \caption{Shape of $\functionNu[{\PDPratio, \Pratio, \loadFactor}]$
    versus $\Pratio$ for some $\PDPratio$'s and $\loadFactor$'s.}
  \label{fig:nu}
\end{figure}

Fig. \ref{fig:nu} shows the shape of
$\functionNu[{\PDPratio, \Pratio, \loadFactor}]$ versus $\Pratio$ for
some values of $\PDPratio$ and $\loadFactor$. As can be
verified, $\functionNu[{\PDPratio, \Pratio, \loadFactor}]$ decreases
as either $\loadFactor$ or $\PDPratio$ increases. This dependency of
$\functionNu[{\PDPratio, \Pratio, \loadFactor}]$ wrt $\loadFactor$ is
justified by the higher resistance to multipath due to increasing 
the length of a single frame \cite{bacci, gezici1}. Similarly to 
$\functionMu[{\PDPratio, \Pratio}]$,
$\functionNu[{\PDPratio, \Pratio, \loadFactor}]$ is a
decreasing function of $\PDPratio$ when $\Pratio$ and $\loadFactor$ are
fixed, since the neglected paths are weaker as $\PDPratio$ increases.
Taking into account the dependency of
$\functionNu[{\PDPratio, \Pratio, \loadFactor}]$ wrt $\Pratio$,
it can be verified that $\functionNu[{\PDPratio, \Pratio, \loadFactor}]$ is
not monotonically decreasing as $\Pratio$ increases. In other words,
an ARake receiver using MRC does not offer the optimum performance in
mitigating the effect of SI, but it is outperformed by PRake receivers
whose $\Pratio$ decreases as $\PDPratio$ increases. This behavior is
due to using MRC, which attempts to gather all the signal energy to
maximize the signal-to-noise ratio (SNR) and substantially ignores the SI.
In this scenario, a minimum mean square error (MMSE) combining criterion, 
while more complex, might give a different comparison.

\section{Analysis of the Nash Equilibrium}\label{sec:performance}

\subsection{Analytical Results}

Using Prop. \ref{th:new} in (\ref{eq:utility}) and
(\ref{eq:minimumPower}), it is straightforward to obtain the utilities
$\utilityStar[k]$ at the Nash equilibrium, which are independent
of the channel realizations of the other users, and of SI:
\begin{align}\label{eq:utilityLSA}
  \utilityStar[k]&\as\hSP[k]\cdot\frac{\infobits}{\totalbits}\rate[]
  \frac{\efficiencyFunction[{\SINRStarInf}]}{\sigma^2\SINRStarInf}\nonumber\\
  &\,\,\times\left(1-\SINRStarInf\cdot\left[(\userno-1)
    \functionMu[{\PDPratio,\Pratio}]+
    \functionNu[{\PDPratio,\Pratio,\loadFactor}]\right]/\gain\right).
\end{align}
Note that (\ref{eq:utilityLSA}) requires the knowledge
of the channel realization for user $k$.
Analogously, (\ref{eq:requirement}) translates into
\be\label{eq:requirementLSA}
  \frameno\ge\left\lceil\SINRStarInf\cdot
  \left[(\userno-1)\functionMu[{\PDPratio,\Pratio}]+
    \functionNu[{\PDPratio,\Pratio,\loadFactor}]\right]/\pulseno\right\rceil,
\ee
where $\lceil\cdot\rceil$ is the ceiling operator. If (\ref{eq:requirementLSA})
does not hold, some users will end up transmitting at maximum power 
$\pmax[]$.

\begin{proposition}\label{th:loss}
In the asymptotic case where the hypotheses of Prop. \ref{th:new} hold,
the loss $\loss=\utilityStar[{k_A}]/\utilityStar[{k}]$ of a PRake receiver 
wrt an ARake receiver in terms of achieved utilities converges a.s. to
\be\label{eq:loss}
  \loss \as\functionMu[{\PDPratio,\Pratio}]\cdot
    \frac{\gain-\SINRStarInf\left[(\userno-1)\functionMuA[\PDPratio]+
        \functionNuA[{\PDPratio,\loadFactor}]\right]}
         {\gain-\SINRStarInf\left[(\userno-1)\functionMu[{\PDPratio,\Pratio}]+
        \functionNu[{\PDPratio,\Pratio,\loadFactor}]\right]},
\ee
where $\utilityStar[{k_A}]$ is the utility achieved by an ARake receiver.
\end{proposition}

The proof of Prop. \ref{th:loss} can be found in \cite{bacci2}.
Equation (\ref{eq:loss}) also provides a system design criterion. Given
\pathno, \pulseno, \frameno, \userno and \PDPratio, a desired loss \loss
can in fact be achieved using the ratio \Pratio obtained by numerically
inverting (\ref{eq:loss}). 

\subsection{Simulation Results}\label{subsec:simulation}
Simulations are performed using the iterative algorithm described 
in \cite{bacci}. We assume that each packet contains $100\,\text{b}$
of information and no overhead (i.e., $\infobits=\totalbits=100$).
We use the efficiency function
$\efficiencyFunction[{\SINR[k]}]=(1-\text{e}^{-\SINR[k]/2})^\totalbits$
as a reasonable approximation to the PSR. Using $\totalbits=100$,
$\SINRStarInf=11.1\,\text{dB}$. We also set $\rate[]=100\,\text{kb/s}$,
$\sigma^2=5 \times 10^{-16}\,\text{W}$, and $\pmax[]=1\,\mu\text{W}$.
To model the UWB scenario, the channel gains are assumed as in
Sect. \ref{sec:interference}, with $\varUser[k]=0.3
\distance[k]^{-2}$, where $\distance[k]$ is the distance between the $k$th
user and the base-station. Distances are assumed to be uniformly distributed
between $3$ and $20\,\text{m}$.

\begin{figure}
  \centering
  \includegraphics[width=8.5cm]{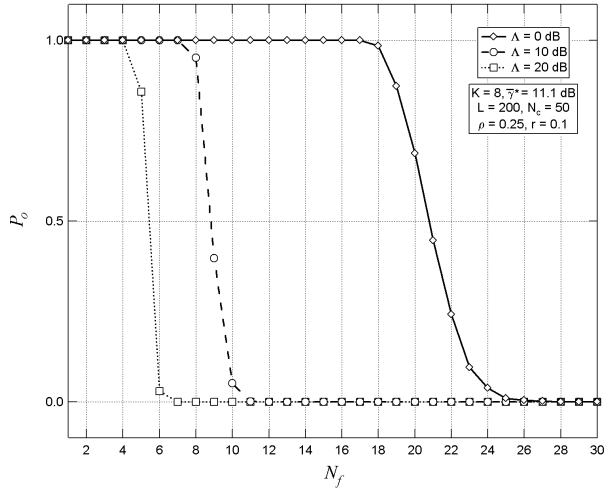}
  \caption{Probability of having at least one user transmitting
    at maximum power versus number of frames.}
  \label{fig:minimumNf}
\end{figure}

Fig. \ref{fig:minimumNf} shows the probability \Po of having at least
one user transmitting at the maximum power, i.e.,
$\Po=\Prob\{\max_k\power[k]=\pmax[]=1\,\mu\text{W}\}$, as a function of
the number of frames \frameno. We consider $10\,000$ realizations of the
channel gains, using a network with $\userno=8$ users, $\pulseno=50$,
$\pathno=200$ (thus $\loadFactor=0.25$), and PRake receivers with
$\prake=20$ coefficients (and thus $\Pratio=0.1$). Note that the slope of \Po
increases as $\PDPratio$ increases. This phenomenon is due to reducing
the effects of neglected path gains as \PDPratio becomes higher, which,
given \frameno, results in having more homogeneous effects of neglected
gains. Using the parameters above in (\ref{eq:requirementLSA}), the minimum
value of \frameno that allows all \userno users to simultaneously achieve the
optimum SINRs is $\frameno=\{21,9,6\}$ for
$\PDPratio=\{0\,\text{dB}, 10\,\text{dB}, 20\,\text{dB}\}$, respectively.
As can be seen, the analytical results closely match with simulations. It is
worth emphasize that (\ref{eq:requirementLSA}) is valid for both \pathno and
\prake going to $\infty$, as stated in Prop. \ref{th:new}. In this example,
$\prake=20$, which does not fulfill this hypothesis.
This explains the slight mismatch between theoretical and simulation
results, especially for small \PDPratio's. However, the authors have found
showing numerical results for a feasible system to be more interesting than
simulating a network with a very high number of PRake coefficients.

\begin{figure}
  \centering
  \includegraphics[width=8.5cm]{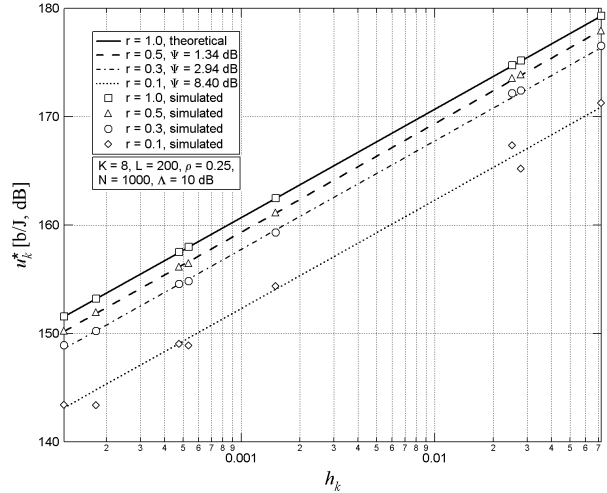}
  \caption{Achieved utility versus channel gain at the Nash equilibrium
    for different ratios $\Pratio$.}
  \label{fig:loss}
\end{figure}

Fig. \ref{fig:loss} shows a comparison between analytical and numerical
achieved utilities versus the channel gains $\channelgain[k]=
\vectornorm{\vecpathgain[k]}^2$. The network parameters are
$\userno=8$, $\pathno=200$, $\pulseno=50$, $\frameno=20$,
$\PDPratio=10\,\text{dB}$, $\loadFactor=0.25$. The markers correspond
to the simulation results given by a single realization of the path gains.
Some values of the receiver coefficients are considered. 
The solid line represents the theoretical achieved utility, computed 
using (\ref{eq:utilityLSA}) with $\Pratio=1$. The
dashed, the dash-dotted and the dotted lines have been obtained by
subtracting from (\ref{eq:utilityLSA}) the loss \loss, computed as in
(\ref{eq:loss}). Using the parameters above, $\loss=\{1.34\,\text{dB},
2.94\,\text{dB}, 8.40\,\text{dB}\}$ for $\Pratio=\{0.5,0.3,0.1\}$,
respectively. It is worth noting that such lines do not consider the
effective values of $\hSP[k]$, as required in
(\ref{eq:utilityLSA}),\footnote{This is also valid for the case ARake,
since $\hSP[k]=\channelgain[k]$.} since they make use of the asymptotic
approximation (\ref{eq:loss}).
The analytical results closely match the actual
performance of the PRake receivers, especially recalling that the results
are not averaged, but only a single random scenario is considered.
As before, the larger the number of $\prake$ coefficients is, the smaller
the difference between theoretical analysis and simulations is.

\section{Conclusion}\label{sec:conclusion}

In this paper, we have used a large system analysis to study performance
of PRake receivers using maximal ratio combining schemes when energy-efficient
PC techniques are adopted. We have considered a wireless data
network in frequency-selective environments, where the user terminals
transmit IR-UWB signals to a common concentration point. Assuming the
averaged power delay profile and the amplitude of the path coefficients to be
exponentially decaying and Rayleigh-distributed, respectively, we have obtained
a general characterization for the terms due to self-interference and 
multiple access interference. The expressions are dependent only on the network
parameters and the number of PRake coefficients. A measure of the loss of
PRake receivers with respect to the ARake receiver has then been proposed
which is completely independent of the channel realizations. This theoretical
approach may also serve as a criterion for network design, since it is
completely described by the network parameters.

\begin{thebibliography}{99}





\bibitem{saraydar2}
C. U. Saraydar, N. B. Mandayam and D. J. Goodman, ``Efficient power
control via pricing in wireless data networks,'' \emph{IEEE Trans. Commun.},
Vol. 50 (2), pp. 291-303, Feb. 2002.



\bibitem{bacci}
G. Bacci, M. Luise, H. V. Poor and A. M. Tulino, ``Energy-efficient power
control in impulse radio UWB wireless networks,'' preprint.
[Online]. Available:
http://arxiv.org/pdf/cs/0701017.

\bibitem{proakis}
J. G. Proakis, \emph{Digital Communications}, 4th ed. New York, NY, 
USA: McGraw-Hill, 2001.



\bibitem{gezici1}
S. Gezici, H. Kobayashi, H. V. Poor and A. F. Molisch, ``Performance
evaluation of impulse radio UWB systems with pulse-based polarity
randomization,'' \emph{IEEE Trans. Signal Process.}, Vol. 53 (7), pp.
2537-2549, Jul. 2005.


\bibitem{schuster}
U. G. Schuster and H. B\"{o}lcskei, ``Ultrawideband channel modeling
on the basis of information-theoretic criteria,'' \emph{IEEE Trans.
Wireless Commun.}, 2007, to appear.


\bibitem{molisch}
A. F. Molisch, J. R. Foerster and M. Pendergrass, ``Channel models for
ultrawideband personal area networks,'' \emph{IEEE Wireless Commun.}, Vol. 10
(6), pp. 14-21, Dec. 2003.

\bibitem{bacci2}
G. Bacci, M. Luise and H. V. Poor, ``Performance of rake receivers in
IR-UWB networks using energy-efficient power control,'' preprint. 
[Online]. Available: http://arxiv.org/pdf/cs/0701034.




\end{thebibliography}
\end{document}